\documentclass{article}
\usepackage{graphicx} 
\usepackage[colorlinks=true, linkcolor=blue, urlcolor=blue, citecolor=blue]{hyperref}
\usepackage[style=authoryear,backend=biber]{biblatex}
\title{Digital Domination: A Case for Republican Liberty in Artificial Intelligence}
\author{Matthew David Hamilton}
\date{published March 20, 2025}

\begin{document}

\maketitle

\noindent{This version is a pre-print of a now published article at \textit{Oxford Intersections: AI in Society}. It can be found at  \href{https://doi.org/10.1093/9780198945215.003.0087}{doi.org/10.1093/9780198945215.003.0087}.}

\begin{abstract}
    Artificial intelligence is set to revolutionize social and political life in unpredictable ways, raising questions about the principles that ought to guide its development and regulation. By examining digital advertising and social media algorithms, this article highlights how artificial intelligence already poses a significant threat to the republican conception of liberty—or freedom from unaccountable power—and thereby highlights the necessity of protecting republican liberty when integrating artificial intelligence into society. At an individual level, these algorithms can subconsciously influence behavior and thought, and those subject to this influence have limited power over the algorithms they engage. At the political level, these algorithms give technology company executives and other foreign parties the power to influence domestic political processes, such as elections; the multinational nature of algorithm-based platforms and the speed with which technology companies innovate make incumbent state institutions ineffective at holding these actors accountable. At both levels, artificial intelligence has thus created a new form of unfreedom: digital domination. By drawing on the works of Quentin Skinner, Philip Pettit, and other republican theorists, this article asserts that individuals must have mechanisms to hold algorithms (and those who develop them) accountable in order to be truly free
\end{abstract}


\section{Introduction}

Since OpenAI introduced ChatGPT, concerns about the ways artificial intelligence (AI) will change the world have exploded. Because AI seems likely to revolutionize important but controversial areas of political life such as policing, redistricting, and war-making \parencite{cho_ai_2022, cumming-bruce_un_2020, lipton_artificial_2023},  popular commentators, academics, lawmakers, and more have understandably sought ethical principles to ensure that AI contributes to, rather than detracts from, human flourishing. Toward this end, scholars have advanced different normative frameworks to guide AI’s development and regulation, promoting and critiquing approaches centered on human rights \parencite{livingston_future_2019, risse_human_2019, dubber_ai_2020}, fairness \parencite{benjamin_race_2019, hoffmann_where_2019, dubber_were_2020, lepri_fair_2018}, virtue ethics \parencite{farina_ai_2024}, and more \parencite{rafanelli_justice_2022, stahl_right_2023, tasioulas_first_2019, zuboff_age_2019}. The prevalence and importance of these conversations has even caught the attention of religious leaders such as Pope Francis, who participated in the G7 summit to convey his concerns about AI to world leaders \parencite{bubola_first_2024}. 

Freedom is a key value in these discussions, but AI’s relationship to freedom is unclear. In many ways, AI promises to increase human freedom: It may, for example, increase free time by automating menial labor or generate new entertainment options tailored to individuals’ unique tastes. At the same time, AI has already violated a deeper and more nuanced kind of freedom called republican liberty. This conception of liberty was both identified and promoted by 20th-century normative theorists like Quentin Skinner and Philip Pettit to sever the theoretical tie between interference and unfreedom. While their contemporary scholarship denied freedom as a lack of interference or obstruction, proponents of republican liberty argued that individuals may be unfree even if they do not experience direct interference in their lives or actions. Moreover, they argued that interference alone is not necessarily unjust.

In this article, I demonstrate how the adoption of AI already  infringes upon this republican conception of liberty, showing how it subjects users to domination—to arbitrary and unaccountable power—in order to make the case that republican liberty offers an indispensable guide for the integration of AI into society. Simply put, AI gives algorithms, and the corporations that develop them, the power to influence individuals and political communities in various ways, whether or not direct or intentional interference actually takes place. While some scholarship has already incorporated republican liberty into broader analyses of AI and freedom, this article seeks to isolate it, to identify republican liberty as a key diagnostic tool for understanding the moral threat AI poses and to concretely show its importance for justly integrating AI into society. 

Toward these ends, I first define republican liberty and distinguish it from alternative conceptions of freedom, like Isaiah Berlin’s  negative and positive definitions, that have been integral to  analyses of AI and freedom. According to republican liberty, protecting freedom requires empowering dominated individuals, especially by giving them mechanisms to hold more powerful parties accountable for their actions. Second, I apply republican liberty to digital advertising and social media algorithms, kinds of AI that people already interface with regularly, to illustrate how AI already infringes upon republican liberty in everyday life.

In doing so, I identify two overlapping spheres of domination: the individual and the political. Individually, these algorithms infringe upon human freedom by shaping individuals’ behavior and thinking even without directly interfering in their lives. Moreover, individuals do not have any means to hold the algorithms, or the algorithms’ developers, accountable for the influence they wield. Politically, algorithms, and the companies that develop them, can influence political processes like elections without appropriate accountability mechanisms. The multinational nature of these corporations and the algorithm-based platforms they share with consumers make it difficult for nation-states to hold accountable bad faith actors who wish to influence a political community from outside it. Similarly, the speed with which these corporations innovate also makes regulating them difficult, exacerbating the accountability deficit even within strong states. After detailing these two sites of ongoing domination, I conclude by sketching out some concrete recommendations for both developing and regulating AI more broadly in ways that respect republican liberty

\section{Republican Liberty}

In its most recent articulation, republican liberty developed in response to Berlin’s distinction between positive and negative liberty, which has become the predominant framework for organizing and understanding conceptions of freedom today. According to Berlin, the modern ideal of liberty tends toward the negative conception, which he defines as being able to “act unobstructed by others” \parencite[p.~169]{berlin_liberty_2002}. “If I am prevented by others from doing what I could otherwise do,” Berlin continues, “I am to that degree unfree,” and this unfreedom can be described as coercion or slavery when it reaches an extreme level \parencite[p.~169]{berlin_liberty_2002}. Berlin calls this negative liberty, for it is the absence of restraints or impediments to action that defines such freedom. 

Positive liberty, in contrast, implies a “wish on the part of the individual to be his own master” \parencite[p.~178]{berlin_liberty_2002}. Proponents of positive liberty generally fear that individuals’ nature, desires, or emotions may rule over them in much the same way that a master rules over a slave, creating a kind of internal or spiritual unfreedom. Accordingly, positive freedom entails the ability to act in particular ways, especially by bringing the ostensibly baser elements of the human psyche under rational control, and it  therefore requires the possession of certain virtues like self-control, rather than simply the absence of obstruction or interference.

Against this framework, a third conception of freedom, primarily called republican liberty, was identified during the second half of the 20th century as proponents of this liberty found that the widespread acceptance of Berlin’s categorization overlooked a key kind of unfreedom. Imagine, for example, a slave with a generally beneficent master who never interferes with the slave’s life but nonetheless possesses the power to do so at any moment. Alternatively, consider the spouse who offers their often belligerent partner excessive praise to avoid their temper \parencites[p.~581]{pettit_freedom_1996}[pp.~22-23]{pettit_republicanism_2000}. In both cases, the slave and the spouse enjoy negative liberty insofar as they do not experience interference, but from a more critical perspective, they suffer significant unfreedom: Even if they live mostly uninterrupted lives, they are both unconditionally subject to another’s jurisdiction. (For arguments favoring non-interference over republican liberty, see \cite{laborde_how_2008, laborde_liberty_2008}.)

Proponents of republican liberty give this condition of unfreedom various names (including subjection to domination, arbitrary power, or alien control), and I will use domination to designate it moving forward. Pettit offers the clearest distillation of domination by breaking the core of the moral claim into three parts: “Someone has such power over another, someone dominates or subjugates another, to the extent that (1) they have the capacity to interfere (2) with impunity and at will (3) in certain choices that the other is in a position to make” \parencite[p.~578]{pettit_freedom_1996}. In other words, domination takes place when a more powerful party can interfere with a weaker party’s life solely at the more powerful party’s discretion—that is, without concern for the weaker party’s desires or opinions—and without accountability for doing so (Pettit, 2000, p. 22). Conversely, individuals enjoy republican liberty when they possess a kind of “antipower” that allows them to exercise reciprocal power over the individuals or institutions that govern their lives \parencite[p.~588]{pettit_freedom_1996}. Republican liberty requires “not just that the doors be open but that there be no doorkeeper who can close a door—or jam it, or conceal it—more or less without cost” \parencite[p.~709]{pettit_instability_2011}.

This republican conception of liberty offers at least two unique insights that normative analysis based on negative or positive freedom overlooks. First, an individual may suffer unfreedom even without experiencing any interference or obstruction, for the power to interfere, rather than actual interference, generates unfreedom. To reuse the master-slave analogy, a slave will likely tailor his actions to t the interests of his master under the reasonable expectation that the master will intervene if he does not; in this way, the master controls the slave’s behavior, rendering the slave unfree, without ever actively interfering with the slave’s life. Freedom, therefore, requires more than going about life unimpeded.

Second, interference should only be considered unjust when it involves “a more or less intentional attempt to worsen an agent’s situation of choice” \parencite[p.~578]{pettit_freedom_1996}. If, for example, I prevent someone from occupying her favorite seat on the Metro by unknowingly taking the seat several stops before she gets on the train, I will interfere with her plans even though I have committed no injustice by doing so. Similarly, interference may be appreciated in some cases as “an alcoholic may thank you for locking up the booze cupboard” even though this limits the alcoholic’s available actions \parencite[p.~108]{laborde_republican_2008}. In both cases, interference limits personal choice, but it does not unjustly infringe upon an individual’s freedom: The first example shows only accidental interference, whereas the second illustrates an improvement to the alcoholic’s condition since it protects his ability to choose sobriety. Republican liberty, therefore, shifts freedom’s relationship with interference, making freedom simultaneously more and less than interference.

The application of this liberty has been widespread. Early proponents of this kind of liberty like Machiavelli and John Locke used this freedom to challenge political power in their day \parencite{halldenius_locke_2003, skinner_machiavelli_1983}. Likewise, many early modern feminists, such as Mary Wollstonecraft, Mary Overton, and Mary Astell, similarly conceptualized freedom, and they extended earlier critiques of arbitrary political power especially to patriarchal power \parencite{broad_women_2014,detlefsen_custom_2016, halldenius_mary_2015, springborg_mary_2005}. More recent theorists have continued to adapt republican liberty for new contexts, repurposing republican liberty to support universal basic income or further placing it in dialogue with feminist theory to rethink self-determination \parencite{pettit_republican_2008, may_two_2005}. Republican liberty has, then, been applied to whatever kinds of domination develop, and the possibility of a digital domination imposed by AI warrants a new application of republican
liberty.

\section{AI and Republican Liberty}

Initially developed for simple calculations, algorithms—or sets of code that instruct computers to perform tasks based on given data—now drive complex yet commonplace AI systems. In digital advertising and social media, algorithms identify patterns and preferences in user data and optimize content recommendations and targeted advertisements to make the experience more targeted and efficient. As these algorithms and their capabilities continue to evolve, they will inevitably become more widespread, influencing practices in healthcare, criminal justice, national defense, and more.

These technological developments and their revolutionary potential led Mathias Risse to issue an “urgent” call for more philosophizing about AI in 2019, especially among human rights scholars \parencite[p.~1]{risse_human_2019}. Risse argued that “the exercise of each human right on the UDHR”—the Universal Declaration of Human Rights—“is affected by technologies, one way or another,” and he surveys many possible moral harms, intentional and otherwise, that could arise from the development and implementation of AI \parencite[p.~11]{risse_human_2019}. Since then, research on the intersection of AI and ethics has begun to blossom, addressing many of the issues Risse identified and going beyond his recommendations as well. Some scholars have explored specific liberties that AI threatens, like freedom of religion or speech \parencite{ashraf_exploring_2022, massaro_siri-ously_2015, risse_human_2019}, while many more have worked to clarify concerns about users’ data (\cite{manheim_artificial_2019, saetra_big_2021}; see also, \cite[p.~12]{risse_human_2019}; \cite[p.~75]{sahebi_social_2022}). More broadly, scholars have also described the threats AI poses to democracy \parencite{christiano_algorithms_2022, manheim_artificial_2019}.

Given this, philosophers and political theorists have unsurprisingly discussed the intersection of AI and different conceptions of freedom. In the Political Philosophy of AI, Mark Coeckelbergh provides a helpful overview for the way that AI interacts with political philosophy, and he devotes a chapter to the relationship between AI and different theories of freedom \parencite{coeckelbergh_political_2022}. Coeckelbergh’s analysis here, however, centers Berlin’s negative and positive liberty, as well as a Marxist conception of liberty, and overlooks republican liberty. Similarly, while Siavosh Sahebi and Paul Formosa have explored social media algorithms’ negative impact on individual well-being, they engage different conceptions of autonomy, a sister concept to republican liberty \parencite{sahebi_social_2022}. 

Some recent scholarship directly incorporates republican liberty, but it typically does so without maintaining a clear distinction between republican liberty and alternative conceptions of freedom or without focusing strictly on AI. Henrik Skaug Saetra, for example, identifies the republican tradition as an influence and incorporates some of its insights throughout his analysis of data privacy and freedom, broadly defined. Saetra provides a helpful starting point for the argument I will make here, but he uses a pluralistic conception of freedom that often employs republican liberty in tandem with other related concepts, like negative liberty, independence, and autonomy, without consistently differentiating between the conceptions (\cite[pp.~15, 29]{saetra_big_2021}; for Saetra's use of republican liberty and related ideals, see pp.~46-48, 59-61, 90, 96). Similarly, Coeckelbergh’s more recent \textit{Why AI Undermines Democracy and What to Do About It} incorporates republican liberty into a broader analysis of AI and democratic values \parencite[pp.~44-42, 101-102]{coeckelbergh_why_2024}.\footnote{Coeckelbergh also mentions a “republican” conception of liberty in \textit{Green Leviathan or the Poetics of Political Liberty}, but he defines this liberty with reference to Plato's \textit{Republic} rather than its traditional association with Quentin Skinner, Philip Pettit, and thinkers inspired by Roman liberty \parencite[p.~22]{coeckelbergh_green_2021}.} Marianna Capasso, in contrast, applies republican liberty to nudging—features intended to influence users’ choices through digital interface design rather than the elimination of alternatives—without focusing on AI or algorithms more specifically \parencite{capasso_manipulation_2022}. In contrast to these approaches, I intend to isolate republican liberty, to identify it especially as a key value for diagnosing and regulating the threats AI poses.

The most thorough discussions of AI and republican liberty come from Jacob Sparks and Athmeya Jayaram’s “Rule by Automation” and Jamie Susskind’s \textit{The Digital Republic}. Like others, Sparks and Jayaram provide a survey of AI’s interaction with different conceptions of freedom, including republican liberty. However, instead of exploring the ways that AI might violate republican liberty, they argue that algorithmic government may protect republican liberty better than human rule does \parencite{sparks_rule_2022}. According to Sparks and Jayaram, algorithmic governance can decrease domination for three reasons. First, algorithms do not have wills or intentionality, so they cannot be considered dominating agents. Second, algorithms eliminate the need for human discretion in applying laws and policies, similarly reducing domination. Third, algorithms eliminate the biases in human judgment that are determined by mood, hunger, pride, and the like \parencite[pp.~205-6]{sparks_rule_2022}.

Since I argue that AI threatens republican liberty, the two arguments seem to clash, but a close reading reveals them to be complementary. Even as they defend algorithmic governance, Sparks and Jayaram recognize that algorithms could exacerbate domination. For example, if a bank uses an algorithm to determine who gets a loan, bank officials must be “under the equal influence of citizens” so that customers can challenge decisions made by the algorithms and ensure that it does not “simply transmit the will of its creators” \parencite[pp.~208]{sparks_rule_2022}. Because of this, they conclude that “rule by automation could reduce domination under the right conditions” \parencite[pp.~208]{sparks_rule_2022}. By clearly identifying the conditions under which AI poses a threat to republican liberty and proposing some remedies, I therefore corroborate Sparks and Jayaram’s argument, develop further some of their nascent regulatory proposals, and demonstrate the broader salience of republican liberty for thinking about AI.

Susskind, meanwhile, offers an extraordinarily accessible overview of the relationship between republican liberty and AI, and he helpfully includes a massive array of morally troubling cases. While the accessibility makes the text both a great introduction to the subject and an easy read, it naturally comes at the cost of some theoretical clarity: Like many of the authors mentioned earlier, Susskind ends up conflating importantly distinct concepts. For example, Susskind argues that “the most basic goal of digital republicanism is the survival of the democratic state itself” \parencite[p.~137]{susskind_digital_2022}. Following this logic, he pushes for algorithmic regulation on democratic, rather than republican, grounds as he argues that algorithms must “be (a) technically sound, and (b) consistent with the moral standards of the community in the context in which they are used, as determined in democratic processes” (\cite[p.~260]{susskind_digital_2022}; for the domination principle, see p. 137). In contrast to this, republican thinkers like Pettit have argued the converse, that democracy’s value comes from its ability to protect republican liberty, and it is easy to imagine how even a democratic community’s “moral standards” may violate some of its members’ republican liberty \parencite[pp.~7-8, 186]{pettit_republicanism_2000}.

Moreover, Susskind intentionally leaves questions about global governance mostly unaddressed \parencite[pp.~13, 203 ff.]{susskind_digital_2022}. While he rightly recognizes the challenges in building cross-cultural cooperation, he minimizes the degree to which national-level regulation abdicates international power to multinational corporations (MNCs). Susskind argues that a national approach to AI regulation creates “an extra burden for transnational technology companies,” but, as I show in this article, nation-centric regulation instead creates tremendous latitude for algorithms, their developers, and their users to act with impunity \parencite[p.~208]{susskind_digital_2022}. Recognizing that “corporations, international manufacturers and other such agencies with potentially global reach” may threaten republican liberty, republican thinkers have emphasized the need to protect states from international domination (\cite[p.~101]{skinner_slogans_2010}; see also, \cite{laborde_republicanism_2010}; \cite{pettit_republican_2010}). Because of this, the multinational questions about AI governance must be considered paramount, as AI cannot otherwise be well-regulated. Therefore, in contrast to the existing literature, I focus my analysis more narrowly on republican liberty to accentuate its unique value, and I emphasize the broader, especially transnational, challenges AI poses to human well-being.

My analysis, then, adds to this scholarship on ethics and AI by proving the value of republican liberty for integrating AI into society in a way that supports humankind. Using digital advertising and social media algorithms as cases, I show how republican liberty helps identify sites of injustice that other approaches may overlook, revealing two levels—the individual and the political—where AI violates freedom today. Importantly, these levels are not entirely discrete but instead mutually reinforce one another to exacerbate domination, and together they demonstrate the need for taking republican liberty seriously when developing, implementing,
and regulating AI.

\section{Individual}

The first level of domination is individual, and the individual level itself has two interconnected, constitutive components. First, individuals change their behavior when they interact with algorithms, whether or not the algorithms actively interfere with action. Here, algorithms shape behavior to benefit their creators without considering the well-being of their users, and users generally do not possess much, if any, reciprocal antipower over and against the algorithms. Second, AI dominates individuals ideationally as it possesses the power to mold thoughts and perceptions of the world. This influence can again take place at the discretion and interest of the algorithms’ creators without consideration for the users’ well-being or antipower.

Starting with individuals’ behavior, algorithms can constrain behavior even without interfering. For example, consider the individual who, surreptitiously shopping for an engagement ring for his partner online, must utilize an incognito browser to avoid triggering incessant targeted advertisements across different devices that would interfere with his plans to surprise his partner. Here, the shopper experiences domination, even if his actions are unobstructed. No one has yet interfered with his ability to shop online, to purchase a present, or to plan the surprise. Nonetheless, like the slave flattering a master to avoid punishment, the shopper alters his behavior due to concerns about the advertising algorithm’s response to his actions. Moreover, the shopper has no reciprocal control over the advertisements that will appear—no ability to prevent the algorithm from ruining the surprise—so he must use an anonymous browser to preemptively avoid the interference. While the stakes in this case are admittedly low, the algorithm still influences and shapes behavior, and an analysis centered on negative liberty would overlook this unfreedom due to the lack of interference.

More substantively, algorithms pose an ideational threat to individuals’ republican liberty. Social media platforms exemplify this domination most obviously, and concerns about the influence of social media algorithms over human thought have recently run rampant. Research facilitated by Facebook demonstrated that manipulating News Feeds toward more or less positive content created social contagion: People shared more positive or negative emotion depending on the positive or negative content fed into their timelines (\cite{kramer_experimental_2014}; for critical engagement with this research, see \cite{meyer_two_2015}). Meanwhile, the Wall Street Journal, which has published extensively on Facebook’s algorithm, revealed that attempts to limit incendiary content only ended up amplifying the controversial posts further, and CEO Mark Zuckerberg rejected proposals that might limit the proliferation of such content due to fears that it would hurt Facebook’s bottom line \parencite{hagey_facebook_2021, horwitz_facebook_2021}.

Of course, this problem is not unique to Facebook. The New York Times has published several pieces that illustrate how YouTube’s algorithm, whose recommendations account for “more than 70\% of all time spent on the site,” has radicalized young men by feeding them extremist content \parencite{roose_making_2019}. And recent debates in the United State over TikTok, which is owned by the Chinese company ByteDance, have reflected similar concerns as lawmakers and much of the American public fear that China could promote or hide content to the detriment of the U.S. population. As a foreign MNC, the American public would have no means to hold ByteDance accountable if an actor chose to use the algorithm maliciously. In both the Facebook and TikTok cases, algorithms subject individuals to digital domination: Algorithms, to return to Pettit’s formula, have “the capacity to interfere . . . with impunity and at will” in the emotional experience and political beliefs of users, influencing them subconsciously through content amplification or suppression \parencite[p.~578]{pettit_freedom_1996}.

This issue also extends beyond an individual’s political beliefs and behavior, as algorithms have shown influence on interpersonal relationships too. Consider, for example, a teenager who tailors her behavior to Snapchat’s friend-ranking algorithm in order to earn a higher friendship ranking on the app. While this feature was ostensibly intended to reflect the character of users’  friendships, it ends up governing, rather than simply describing, these friendships and incentivizes users to act in ways that t the algorithm \parencite{jargon_snapchats_2024}. Ironically, academics experience something similar with journal metrics, like Impact Factor, that were developed to quantify the most prominent journals but may now shape scholars’ behavior by encouraging them to make their research more amenable to highly rated journals. While both of these indices were developed for descriptive purposes, individuals who interface with them end up tailoring their behavior to the algorithm in ways reminiscent of the slave stroking a master’s ego.

Recent research dispels at least some of the fears about social media algorithms’ negative impact \parencite{haroon_auditing_2023, hosseinmardi_causally_2024}, but whether or not these algorithms actually radicalize users, their potential to influence society nonetheless compromises republican liberty. Such algorithms might not be actively interfering with users’ ideas or actions yet, but they nonetheless have the power to do so irrespective of concerns for users’ well-being and, as the TikTok case shows especially well, without any direct accountability. Instead, as illustrated by Zuckerberg above and by X owner Elon Musk below, executives at various MNCs direct the way their companies’ algorithms shape individuals, and these executives are largely insulated from any
widespread accountability.

Now, some may push back against my argument here, pointing out that this problem is neither new nor unique to AI: Television and earlier forms of media have the same ability to manipulate action or perception, so news institutions like the New York Times or Fox News, according to my argument, have long posed a similar threat to republican liberty. I have no disagreement here: Republican liberty has been used to oppose arbitrary power in different contexts, and criticisms of mass media and other institutions on the basis of republican liberty could certainly be developed. These concerns, however, stand outside the scope of this article, as I instead only translate republican liberty to a new and emerging digital context. While AI, due to its sophistication, may pose a greater threat to republican liberty than previous institutions did, my criticism of AI’s dominative capacity is not intended to minimize other violations.

Similarly, some may argue that users have plenty of antipower in these cases, for they can choose not to use algorithm-based platforms, especially by deleting social media accounts. But this counter has two problems. First, algorithms, and the data-gathering operations associated with them, have become so widespread that individuals cannot opt out of interacting with them without facing extreme costs. For example, Google collects personal data through basic digital infrastructure, like emails and cloud-based document storage, so avoiding targeted advertisements based on this information would require almost entirely disconnecting from the internet, even for employment. Second, widespread social media use will likely influence the actions and thoughts of a given political community, so individuals are likely to be subject to arbitrary algorithmic power via the politicians elected and laws enacted, even if they themselves leave these platforms. Individuals, then, cannot escape this domination through their own actions alone. For these two reasons, simply avoiding algorithms does not offer individuals meaningful antipower.

Importantly, many technology companies have begun to offer users some antipower over and against these algorithms. For example, many social media platforms now allow users to select a “Not Interested” option to discourage identified kinds of content from appearing on their timelines. Similarly, Apple has given its users the ability to “Ask App Not to Track” when an application seeks to track a user’s behavior beyond their use of the specific platform. The conclusion below contains a larger discussion of ways to alleviate AI domination, but accountability mechanisms such as these should be recognized, as they directly respond to algorithmic power and create antipower for users.

In any case, analyzing these examples through the lens of republican liberty reveals how algorithms both possess and have already exercised arbitrary power over individuals. Whether or not they and their developers weaponize the power they possess, algorithms have the ability to intervene and shape both behavior and ideation at any time. Users, meanwhile, do not have a means to exercise accountability over and against the corporations that wield these algorithms, and the corporations can act without considering the individual’s best interests. While the cases highlighted here may be relatively harmless when compared to the future injustices AI might perpetrate, their banality nonetheless thereby draws attention to the importance of republican liberty for identifying and fending off the more insidious threats AI poses to human well-being.

\section{Political}

The second level of domination is political, and it also has two components. First, the multinational nature of the technology companies that develop these algorithms and their platforms opens space for international actors to influence other political communities without accountability. The previously mentioned concerns over ByteDance manipulating TikTok’s algorithm illustrates this issue within the United States, but U.S. technology companies, like Meta (Facebook’s parent company) and X, have found themselves embroiled in other states’ domestic political conflicts. Because the technology companies that develop and popularize these algorithms are multinational, they cannot be sufficiently held accountable by any national polity, and they instead open space for outside actors to influence political communities with impunity. Second, the speed and secrecy with which these private companies innovate shield them from robust government oversight and effectively render them self-regulating. Both components illustrate that algorithms already exercise a kind of political domination since they possess the power to intervene in political life without popular accountability. 

On the former point, some American MNCs have become key decision-makers in thorny political situations because of the algorithm-based products they provide to users. For example, the ruling Myanmar military used Facebook to incite hatred and genocide against the Rohingya, a Muslim minority in Myanmar, and the platform had to decide whether to allow or block the government’s content \parencite{mozur_genocide_2018, stevenson_facebook_2018}. Similarly, Musk complied with Turkish political leadership’s demands to censor some content on X leading up to an important election \parencite{stein_twitter_2023}. Since their platforms’ algorithms function as a kind of digital telecommunication infrastructure, MNC leadership inherits the power to police speech when a political community embraces their product.

In both cases, the technology companies allowed their algorithms to become weapons for autocratic governments, and the victims in Myanmar and Turkey had no means to exercise authority over the decisions that executives at Facebook or X made for their political community. While a normative framework centered on negative liberty offers unclear conclusions—the corporations interfered with some individuals’ ability to protest their governments but did not interfere with the state governments’ actions—republican liberty identities the power these technology executives possess as itself morally problematic. In effect, the mismatch between multinational and national parties shields technology MNCs from accountability and gives them arbitrary power over political communities.

More poignantly, the algorithmically guided platforms provide opportunities for outsiders to drive political outcomes. Cambridge Analytica, for example, utilized Facebook data to target advertisements and sway election results in the United States and Great Britain \parencite{rosenberg_how_2018}. Russian hackers have similarly utilized X’s algorithm to foment polarization within the United States \parencite{linvill_that_2019}. Here, the transnational nature of the algorithms’ user base facilitates international domination, for actors can exercise political influence without direct repercussions or accountability. In all of these cases, individuals are subject to domination since, once again, the corporations have “the capacity to interfere” with political communities’ choices “with impunity and at will” \parencite[p.~578]{pettit_freedom_1996}.

At the same time, the second point—the speed of innovation—illustrates the difficulty governments face in keeping up with technology companies. Recognizing this, technology companies have made attempts at self-regulation. For example, when an individual named Victor Miller used ChatGPT to create a chatbot mayoral candidate, OpenAI blocked Miller’s access to the service to end the mayoral bid \parencite{feiger_ai_2024, kelly_ai_2024}. While prohibiting an AI-based candidate to run for office may seem like a reasonable move now, this nonetheless shows OpenAI making a decision that removes a potential candidate from an election ballot. Similarly, Facebook developed what some called a “Supreme Court” to help it regulate its own algorithm and determine the limits of acceptable speech on the platform \parencite{klonick_inside_2021, ovide_facebook_2021}. Again, the decisions made inevitably impact Facebook users who have limited means to exercise reciprocal antipower over the algorithm and its developers. Both cases demonstrate how the lagging nature of government legislation pushes technology companies to self-regulate. This, in turn, exposes individuals to domination as the corporations that develop algorithms are free from government oversight.

As at the individual level, it is worth noting a few accountability mechanisms that have already been developed to curb some of this domination. The forced sale of TikTok ostensibly serves this end in the United States as it makes the algorithm’s influence subject to the U.S. legislators. Similarly, the European Union has enacted both the Digital Markets Act and the Digital Services Act to lessen the domination that technology companies perpetrate against its members by decreasing the ability of these companies to accumulate user data and requiring that they take additional steps to police content on their platforms \parencite{ocarroll_how_2023, satariano_eu_2022}. In doing so, they limit the data-based power that technology companies can accumulate and subject these corporations to more popular government. Just as the technological developments have emerged to grant users more behavioral antipower against the algorithms, these shifts similarly work to increase political antipower, to protect individuals from political domination by the algorithms and the companies that develop them.

Nonetheless, AI threatens to exacerbate political domination since it is difficult to hold technology MNCs accountable within the current nation-state system. Oppressive regimes and outside parties have used the algorithms to propagate their message, sway public opinion, or silence opposition within states, and affected individuals have no means to hold executives at these distant corporations or other malicious agents accountable for their political influence. Moreover, the speed of innovation exacerbates the difficulty states have in regulating AI, placing individuals under even greater domination. These features make it extraordinarily difficult for citizens of any state to possess commensurate antipower over and against the technology companies that develop and share these algorithms with the world.

\section{Conclusion}

In his message on AI, Pope Francis pointed toward the “need to strengthen or, if necessary, to establish bodies charged with examining the ethical issues arising in this field and protecting the rights of those who employ forms of artificial intelligence or are affected by them” \parencite{francis_message_2024}. This essay works toward these ends, identifying widespread yet inchoate unease about algorithms in society as legitimate moral concern about threats to individuals’ republican liberty. Moreover, by using republican liberty in normative analysis, I reveal key sites of injustice that demand attention. At the individual and the political levels, algorithms, and their associated companies, possess the ability to interfere with individuals’ action and thought without considering their well-being or being subject to reciprocal power exercised by the subject individuals. While other analyses based on different conceptions of freedom can offer important insight, republican liberty captures the ways that even unexercised power can cause unfreedom.

More broadly, the ubiquitous adoption of social media and digital advertising algorithms illustrates republican liberty’s value viz-a-viz AI: While AI promises to revolutionize more controversial areas of society than those covered here, these relatively banal examples capitalize upon preexisting concerns and familiarity with algorithms to demonstrate the need to protect republican liberty moving forward. In demonstrating the ways AI dominates us today, I do not seek to prevent the development of AI or to discourage its use in social and political life; instead, I only want to identify a guiding light for developers and regulators, an ideal that can help inspire a domination-free relationship between humans and AI in the future.

In his treatise on antipower, Pettit details three ways that domination may be mitigated: “We may compensate for imbalances by giving the powerless protection against the resources of the powerful, by regulating the use that the powerful make of their resources, and by giving the powerless new, empowering resources of their own” \parencite[pp.~589-90]{pettit_freedom_1996}. Based on these comments, I want to conclude by highlighting a few concrete ways to protect republican liberty moving forward. In simplest terms, developers, regulators, and other interested parties should work to increase individual’s antipower, their ability to hold algorithms and their creators accountable for algorithms’ potential influence. In doing so, individuals should obtain a “degree of counter-control over” algorithms and technology companies, thereby lessening their domination \parencite[p.~108]{laborde_how_2008}.

Accountability can be created in at least two ways. First, developers and the private sector should work to give users more digital power over the algorithms with which they interact. As discussed under individual domination, many corporations have taken steps in this direction already, and they should be further encouraged to do so. Increasing individuals’ ability to prevent apps from tracking personal information across devices serves as one way to do this, but users should also have access to more information about and intentional control over the algorithms that cultivate their unique experiences. Algorithmically determined recommendations should include a viewable explanation for suggestions to better involve users in the decision-making process, and users should have greater ease in overriding or changing algorithms’ suggestions. As Apple and other companies have shown, customers often prefer more control over their data, creating a market-based incentive for lessening domination.

More importantly, however, protecting republican liberty requires creating political institutions to hold algorithms and those who develop them accountable in the corporeal world. Here, Sparks and Jayaram’s argument provides some concrete guidance: AI should be “implemented”—or at least regulated—“by the people’s representatives,” and its processes should be “contestable by citizens, whether by appeal to another automated system or to a human authority” \parencite[p.~208]{sparks_rule_2022}. Given the multinational scope of AI’s influence, the “people” must be reconceptualized to cross national lines. Accordingly, individuals must begin to push for supranational institutions like the EU that can resolve disputes generated by human interactions with algorithms or their developers and that can better hold transnational and multinational actors accountable for the ways they might manipulate political life through algorithms. Without corresponding supranational architecture, domestic political institutions will be largely impotent against the most significant violations AI might perpetrate.

AI will continue to grow its influence on society, and some states have already begun to extend it to more controversial policies, like policing or sentencing. These uses threaten to exacerbate the domination that marginalized groups like women and ethnic and racial minorities already face. While algorithmic optimization of infrastructure like bus routes or sewer systems may gain widespread support, the more extreme instances and possibilities of AI deployment reveal the need for principles that can guide the ongoing integration of AI into society. The cases covered here serve to illustrate that republican liberty must be one of these guides. Individuals must recognize their unease about algorithms’ influence as an important concern for their own republican liberty, and policymakers should respond to this by building supranational institutions that protect this good. If it is developed and integrated with republican liberty in mind, AI may contribute to human society without eroding its most basic freedom.

\printbibliography

\end{document}